\documentstyle[aps,epsfig,floats]{revtex}
\newcommand {\be} {\begin{equation}} 
\newcommand {\ee} {\end{equation}} 
\newcommand {\Be} {\begin{eqnarray*}}
\newcommand {\Ee} {\end{eqnarray*}}
\newcommand {\bey}{\begin{eqnarray}} 
\newcommand {\eey}{\end{eqnarray}} 
\newcommand {\oti}{\tilde\omega} 

\begin{document}

\begin{center}

\Large{\bf Relaxation of classical many-body 
hamiltonians in one dimension}\\

\vspace{0.3cm}

{\large Stefano Lepri}\\

\vspace{0.1cm} {\small\it
Max-Planck-Institut f\"ur Physik komplexer Systeme,
N\"othnitzer Stra{\ss}e 38, D-01187 Dresden, Germany\\
{\rm lepri@mpipks-dresden.mpg.de}\\
}

\end{center}

\date{\today}
\begin{abstract}
\noindent{The relaxation of Fourier modes of hamiltonian chains close to 
equilibrium is studied in the framework of a simple mode-coupling theory. 
Explicit estimates of the dependence of relevant time scales on the energy
density (or temperature) and on the wavenumber of the initial excitation are 
given. They are in agreement with previous numerical findings on the
approach to equilibrium and turn out to be also useful in the 
qualitative interpretation of them. The theory is compared with molecular 
dynamics results in the case of the quartic Fermi-Pasta-Ulam potential. 
}

\vspace{0.2cm}
\noindent{\sl Keywords}: Relaxation to equilibrium, Fourier modes, 
Fermi-Pasta-Ulam model\\
\noindent{\sl PACS numbers}: 05.70.Ln, 05.45.+b, 63.10.+a\\
\noindent{\sf To appear in Phys. Rev. E (1998)}\\
\end{abstract}
\vspace{0.2cm}

\section{Introduction and motivations}

Consider the conceptual experiment where one of the normal modes of 
an idealized, one-dimensional crystal is excited by means of some external source, 
in such a way that the system is brought far from thermal equilibrium. After
switching off the external perturbation, it will relax again to the equipartition
state, described by the canonical measure. The classical question is: how long 
will it take? Such a basic issue was seriously reconsidered in the light of recent 
results of contemporary nonlinear dynamics, in particular after the discovery that 
weakly nonlinear systems may display extremely long relaxation times. The latter
are related to an effective ``freezing'' of some degrees of freedom, i.e. to the 
slow diffusion in phase space (see for example Ref.~\cite{benettin} for a 
recent critical discussion and further bibliographical references). 
Because of their simplicity, hamiltonian chains of oscillators are suitable 
model systems to discuss the problem, both from the
analytical than from the simulation point of view.

Although some theoretical work \cite{tsaur,shepe} indicates that equipartition 
in a strict sense is always attained in the thermodynamic limit for chains at 
finite temperature (or energy per particle), no explicit estimates of the time
scales needed are known. The problem has been attached mainly from the numerical 
side, by looking at the time relaxation of suitable indicators of 
equipartition among the Fourier modes of the chain \cite{pettini,kantz,ruffo,parisi}. 
In particular, recent studies \cite{pettini,thierry} focused on the dependence of 
relaxation times on the energy density or temperature, and some empirical scaling 
laws were found. Nonetheless, no quantitative explanation of the latter is
sofar given. The need for some analytic clue is even more evident as one 
consider that the computational limits of simulations can be rapidly 
reached. This is of course a particularly serious limitation at very low 
energies, when the interaction among modes and the resulting dissipative 
effects are extremely weak.

A further motivation comes from the closely related problem of energy transport 
in such systems. As is well known, the relaxation of fluctuations is strictly
connected to transport coefficients, and the existence of slow time scales must
be reflected somehow in their properties. Indeed, extensive molecular dynamics 
studies \cite{noi} gave evidence of the divergence of the thermal conductivity 
associated to the algebraic decay of the correlation function appearing in the 
corresponding Green-Kubo formula.

The present work aims to give a contribution to the comprehension of both 
questions by studying the relaxation of chains with an acoustic spectrum.
The models we will refer to are introduced Sec. II. When the system is not 
too far from equilibrium, one can rely on well-established theories, like 
the perturbative and mode-coupling approximations that are reviewed in Sect. III. 
Their validity will be compared with the outcomes of numerical simulations 
for the specific case of a interatomic potential with quartic nonlinearities 
(Sect. IV). Besides the original motivations, the comparison
constitutes a nice test of mode-coupling theory and of its typical 
features i.e. the existence of long time-tails with nontrivial exponents, and the 
nonanalytic behaviour of the spectrum of relaxation times. This latter property
is actually a peculiarity of nonequilibrium dynamics in one dimension.
As it will be hopefully clear, this approach will be extremely useful for 
the problems at hand, as they provide estimates of the relaxation 
times of the Fourier modes of the chain. Rather surprisingly, a comparison with 
some previous numerical results will show that the latter are largely the main
cause of slow relaxation to equipartition (Sect. V). 

\section{One-dimensional many-body hamiltonians}

We consider a chain of $N$ anharmonic oscillators and denote with $q_l$ the 
displacement of the $l$-th particle from its equilibrium position. The 
hamiltonian reads as
is  
\be
H = \sum_{l=1}^N \left[{p_l\over 2}^2+V(q_{l+1}-q_{l})\right] \quad;
\label{hami}
\ee
where the usual Born-Von Karman boundary conditions $q_l=q_{l+N}$ are assumed 
and the potential energy is of the form
\be
V(x)\;=\;{x^2 \over 2} + V_A(x) \quad,
\ee
with $V_A$ denoting the anharmonic part of it. 
We will consider homogeneous lattices, so that all the masses are set equal 
to unity and $p_l=\dot q_l$. The lattice spacing is also set to unity as well as
the harmonic frequency, so that all variables in the following are adimensional.
This also means that the sound velocity, as defined in the harmonic approximation, 
is equal to one. 

In the following we will always refer to the energy density $\varepsilon$ (energy
per particle) and/or to the corresponding temperature $k_BT=1/\beta$. Obviously, 
for a strongly nonlinear systems, the two quantities are not in general strictly 
proportional.

The (complex) amplitudes $Q_k$ of the Fourier modes are defined through the 
usual transformation 
\be
Q_k={1\over\sqrt{N}}\sum_{l=1}^N q_l \, e^{i{2\pi k\over N}l} \quad, \qquad
Q_{-k}=Q_k^*, \qquad k=-{N\over2}+1,\ldots, {N\over2} \quad .
\label{modi}
\ee
Once the hamiltonian is expressed in these new canonical variables, the equations of 
motion become
\bey
&&\dot Q_k \;=\; {\partial H \over \partial P_k^*}=P_k \\
&&\dot P_k \;=\; -{\partial H \over \partial Q_k^*} =
-\omega_k^2 Q_k + {\cal F}_k
\label{newton}
\eey
with ${\cal F}_k$ being the interaction force among modes, and we have 
introduced the usual normal-mode frequencies
\be
\omega_k \;=\; 2 \big| \sin\left({\pi k\over N}\right)\big| \quad.
\label{barefreq}
\ee

\section{Estimate of the relaxation times}

The formulation of stochastic equations for the dynamics of the relevant 
variables is rather customary to describe the relaxation close to equilibrium 
\cite{kubo}. The idea is to describe the effective motion of suitable ``slow'' 
observables by reducing the level of description. The general strategy involves 
projection on their subspace, and results in linear non-markovian
equations. Whenever a sharp separation of time scales is possible the latter 
reduce to their markovian limit. The memory term determines the relaxation 
properties and can be estimated self-consistently. Our aim is to apply the 
above procedure to the present system. In the following two subsections we 
summarize the relevant steps.

\subsection{General setting}

Due to the conservation law of total momentum, we expect that in the present
case the slow dynamics should be associated with the long-wavelength Fourier
modes $Q_k$ with $|k|\ll N/2$. Moreover, translational invariance implies that 
each mode is uncorrelated from the others so that we can that consider each 
mode separately. 
\footnote{One can easily convince hymself of this statement by computing for 
instance the correlation $\langle Q_k(t)Q_{k'}^*(0)\rangle$ and imposing that 
$\langle q_l(t)q_{l'}(0)\rangle$ depends only on $|l-l'|$.}
Accordingly let us consider the set of $Q_k$ and $P_k$ as relevant variables
and define the projection operator $\hat {\cal P}$ acting on the scalar observable 
$X$ as
\be
\hat{\cal P}X \;=\;
=\sum _k \left[{\langle X Q_k^*\rangle \over \langle |Q_k|^2\rangle} \,Q_k \,+\, 
{\langle X P_k^*\rangle \over \langle |P_k|^2 \rangle } \, P_k \right]
\ee
The projection of the equation of motion leads to \cite{kubo}
\bey
&&\dot Q_k = P_k \\
&&\dot P_k = -\oti_k^2 Q_k -\int_0 ^t \Gamma_k(t-s) P_k(s) ds  + R_k 
\label{langevin}
\eey
where $R_k= (1-\hat{\cal P}) \dot P_k$ is the so-called random force which is 
related to the memory function by the fluctuation-dissipation theorem
\be
\Gamma_k(t) \;=\; \beta \langle R_k(t)R_k^*(0)\rangle \quad .
\label{fludis}
\ee
and the renormalized frequencies are given, for a generic hamiltonian like 
(\ref{hami}), by 
\be
\oti_k^2 = {1\over \beta \langle |Q_k|^2\rangle}
= (1+\alpha)\, \omega_k^2 \quad ,\qquad
\alpha(\beta) = {1\over \beta} {\int e^{-\beta V(x)} dx
\over \int x^2 \; e^{-\beta V(x)} dx }\;-\;1\quad .
\label{alfa}
\ee
Obviously in the harmonic limit $\alpha \to 0$, and the usual bare dispersion 
relation (\ref{barefreq}) is recovered. Here and in the following we will 
always deal with bounding potentials so that the integrals in (\ref{alfa})
are convergent. The definition (\ref{alfa}) amounts 
then to a renormalization of the sound speed from unity to the temperature
dependent value $\tilde v = \sqrt{1+\alpha}$.

The main object of study will be the normalized correlation function
\be
{\cal G}_k(t) \;=\; \beta \oti_k^2\, \langle Q_k(t)Q_{k}^*(0)\rangle \quad,
\ee
which is defined in such a way that ${\cal G}_k(0)=1$.  It satisfies the 
equation of motion \cite{kubo}
\be
\ddot {\cal G}_k + \oti_k^2 {\cal G}_k \; = \; 
-\int_0 ^t \Gamma_k(t-s) \, \dot {\cal G}_k(s) ds 
\ee
Introducing the Laplace transforms ${\cal G}_k(z)$ and $\Gamma_k(z)$ with
the definition
\be
\Gamma_k(z)\;=\;\int_0^\infty e^{-izt} \Gamma_k(t) dt 
\ee
one has that (with $\dot{\cal G}_k(0)=0$)
\be
{\cal G}_k(z) \;=\; 
{iz+\Gamma_k(z) \over z^2 - \oti_k^2 -iz \Gamma_k(z)} \, 
\label{laplag}
\ee
If the dissipation is small enough with respect to $\oti_k$, the transform 
${\cal G}_k(z)$ has two poles close to the real axis in the complex plane 
approximatively given by
\be
\pm \oti_k -{i\over 2} \, \lim_{z\to \oti_k+i0^+} \, \Gamma_k(z) ,
\label{poles}
\ee
(at fixed wavenumber and provided that the limit exists).
This corresponds to both a shift of the renormalized frequencies and a small 
damping $\gamma_k$, given by the imaginary part of (\ref{poles}) (i.e. by
the real part of $\Gamma_k(\oti_k)$). Accordingly, the inverse of the latter 
defines a characteristic relaxation time of each Fourier mode. 

\subsection{The mode-coupling approximation}

The above results are more or less formal manipulations: we obviously need 
to compute explicitely the memory kernel and the relaxation rates $\gamma_k$ 
by resorting to some approximations.   
A first conceptual difficulty of the projection approach is the fact that 
$R_k$ does not evolve with the full Liouvillean operator associated with $H$ 
\cite{kubo}. One generally bypass the problem by simply replacing 
\be
\langle R_k(t)R_k^*(0)\rangle 
\;\to\;
\langle {\cal F}_k(t){\cal F}_k^*(0)\rangle
\label{replace}
\ee
where the last average is on the full Gibbs measure. In such a way, it is also
implicitely assumed that slow terms possibly contained in ${\cal F}_k$ are 
negligible in the thermodynamic limit (see below).
A second simplification amounts to factorize multiple correlations so that 
the resulting approximate expression of the memory kernel $\Gamma_k(z)$, together
with Eq. (\ref{laplag}), constitute a closed system of equations for ${\cal G}_k$. 
The latter has to be solved self-consistently.

Let us focus on the Fermi-Pasta-Ulam (FPU) potential
\be
V_A(x) \;=\; {1\over 3}g_3 \, x^3 \,+\, {1\over 4}g_4 \, x^4 
\label{fpu}
\ee
so that ${\cal F}_k = {\cal F}_k^{(3)}+{\cal F}_k^{(4)}$ with
\bey
&&{\cal F}_k^{(3)}=
-g_3\omega_{k}\,{1\over \sqrt{N}}\sum_{k_1+k_2=k}
\; \omega_{k_1}\omega_{k_2} Q_{k_1}Q_{k_2}  \\
&&{\cal F}_k^{(4)}=
-g_4\omega_{k}\,{1\over N}\sum_{k_1+k_2+k_3=k}
\omega_{k_1}\omega_{k_2}\omega_{k_3}
Q_{k_1}Q_{k_2}Q_{k_3} \quad ,
\eey
and where the condition on the indexes of the sum is intended to be 
modulo $N$ (quasi-momentum conservation). Obviously, this 
represent a reasonable approximation of a generic anharmonic potential 
in the limit of small anharmonicity.
In the approximation where multiple correlations factorize in products, 
the memory kernel will be a sum of two terms (averages with an odd number 
of $Q$ vanish):
\be
\Gamma_k(t) \;=\;
\beta [ \langle {\cal F}^{(3)}_k(t){\cal F}_k^{(3)*}(0)\rangle
+ \langle {\cal F}^{(4)}_k(t){\cal F}_k^{(4)*}(0)\rangle ]\quad.
\ee
and one can readily evaluate the two contribution from the cubic and quartic 
terms respectively
\bey 
&&\beta \langle {\cal F}_k^{(3)}(t){\cal F}_k^{(3)*}(0) \rangle \;\approx\;
C_3\omega_k^2 \, {1\over N} \sum_{k_1+k_2= k}
{\cal G}_{k_1}(t) \; {\cal G}_{k_2}(t) \;
\label{corfo3}
\\
&&\beta\langle {\cal F}_k^{(4)}(t){\cal F}_k^{(4)*}(0) \rangle \;\approx\;
C_4\omega_k^2 \, {1\over N^2} \sum_{k_1+k_2+k_3 = k}
{\cal G}_{k_1}(t) \; {\cal G}_{k_2}(t) \; {\cal G}_{k_3}(t) \;
\label{corfo4}
\eey
where we have defined the two constants
\be
C_3 \;=\; 3 \,{g_3^2\over \beta (1+\alpha)^2} \quad,\qquad 
C_4 \;=\; 15 \,{g_4^2\over \beta^2 (1+\alpha)^3} 
\label{c3c4}
\ee
The numerical factors come from the counting all the possible factorizations.
Before going further, notice that in the present example we can easily understand
the assumption of neglecting slow components in (\ref{replace}). The force 
${\cal F}^{(4)}_k$ contains indeed a term proportional to $Q_k|Q_k|^2/N$, which
is clearly as slow as $Q_k$ itself. As assumed, its weight vanishes like $1/N$, but
for any finite $N$, it may be then regarded as one of the sources of finite-size 
effects in the numerical simulations.

Even with all the above simplifications the theory remains too complicated to
be solved. A further assumption is then generally required, namely that the
sums in Eqs. (\ref{corfo3}) and (\ref{corfo4}) can be replaced with their values 
at $k=0$ \cite{pomeau}. For the cubic case this amounts to set $k_1=-k_2=k'$ in 
Eq. (\ref{corfo3}), yielding
\be 
\beta\langle {\cal F}_k^{(3)}(t){\cal F}_k^{(3)*}(0) \rangle \; \approx \;
C_3 \omega_k^2\, {1\over N} \sum_{k'} {\cal G}_{k'}^2(t)
\label{appro}
\ee
For the quartic term we can also extend the summation only to the small $k$ 
terms that are almost-resonating, e.g. those for which $k_1-k_2-k_3\;\approx\;0$.
This simplification is justified as in the long-time limit only the slowly
oscillating contributions should be significative.
One can convice hymself that this kind of approximation leads to the same result
(\ref{appro}) also for the quartic term, with $C_3$ replaced by $C_4$.
Finally, in the limit $N\to \infty$, we let $2\pi k/N \,\to\, q$ and replace
the sums with integrals. The above hypotesis that we can set $k=0$ in 
Eqs. (\ref{corfo3}) and (\ref{corfo4}) amounts therefore to say that the memory 
kernel can be written in the form $\Gamma(q,z)=\nu(z)\, q^2$ for $q\to 0$ \cite{pomeau}. 
As a result, one gets the self-consistency relation 
for $\nu$ from the Laplace transform of Eq. (\ref{fludis})
\be
\nu (z) \;\approx\; C \int_0^\infty dt \, e^{-izt}\;
\int {dq \over 2\pi} \, {\cal G}^2(q,t) \quad ,
\label{selfco}
\ee
where $C=C_3+C_4$. This is readily solved by dimensional arguments \cite{pomeau}, 
and yields 
\be
\nu (z) \;\propto\; {C \over \sqrt{\nu(z)\, z}} \quad .
\ee
This last relation implies that $\Gamma(q,z)\,\sim\, z^{-1/3}q^2$, so that
the limit in Eq.~(\ref{poles}) leads to
a non-analytic dependence of the relaxation rates for small wavenumber:
\be
\gamma(q) \;\propto\; \left({C^2 \over \tilde v}\right)
^{1/3} \, q^{5/3} \quad.
\label{gammamct}
\ee
Generally speaking, the behaviour of the relaxation rates with the temperature 
will depend on the specific form of the anharmonic potential. Nevertheless, the 
$q-$dependence should be the same for all one-dimensional models where the 
theory applies. 

\subsection{Kinetic vs. hydrodynamic relaxation}

The above self-consistent result is expected to hold for strong anharmonicity and
on very long (``hydrodynamic'') time scales. On the other hand, in the usual 
perturbative limit, the system is basically a set of weakly interacting harmonic 
oscillators with renormalized frequencies. We can then use the perturbation theory 
in our simple mode-coupling scheme. 
This basically amounts to neglect the dissipation on the r.h.s. of Eq. (\ref{selfco}) 
and approximate ${\cal G}(q,t) \approx \cos \oti(q) t $. In this limit the 
factorization of correlations becomes exact. Accordingly, Eq. (\ref{selfco}) reduces 
then to a simplified version of the usual perturbative formula \cite{pert}, where 
only mode-mode contributions are taken into account. It then follows that
the spectrum of the dissipation rates is proportional to $\omega_k^2$ times the 
sum of two 
terms whose magnitude is given by the constants $C_3$ and $C_4$ respectively. 
In the limit of low temperatures and/or weak couplings the latter scale as
\bey
& \,{(g_3 q)^2\over \beta (1+\alpha)^{5/2}} \;\approx\; (g_3 q)^2 \,\varepsilon &
\qquad ({\rm from \; cubic \; term}) \nonumber\\
& {(g_4 q)^2\over \beta^2 (1+\alpha)^{7/2}} \;\approx\; (g_4 q)^2\, \varepsilon^2&
\qquad ({\rm from \; quartic \; term})
\label{qterm}
\eey
for small wavenumbers.
It is of basic importance to compare this latter (``kinetic'') time scale with the
one determined in the previous subsection. For example, in the case of the quartic FPU 
model we have from Eqs. (\ref{gammamct}) and (\ref{qterm})
\be
{\tau_{hydro} \over \tau_{kin}} \;\sim \; 
q^{1/3} (g_4 \varepsilon)^{2/3} \ll 1 \quad.
\ee
This implies that, for small enough $g_4 \varepsilon$, the initial relaxation 
stage is dominated by the kinetic time scale, up to some (possibly large) crossover 
time where the self-consistency effects become relevant. In Section V we will come
back on this issue to show its importance for the approach to equilibrium.

\section{Molecular dynamics results}

As many assumptions are required in the theory, it is important to compare it
with the outcomes of numerical simulations. To this aim, we considered the quartic 
FPU case ($g_3=0$). In this case, the only relevant parameter is $g_4\varepsilon$
and $g_4\varepsilon\ll 1$ correspond to the weakly chaotic regime \cite{pettini}.

The numerical simulations were performed at constant energy by integrating the 
equations of motion with a third order symplectic algorithm \cite{algor}. 
We generally consider the case where the second constant of motion $P_0=\dot Q_0$ 
is identically zero (no uniform rotations of the chain). Equilibrium 
initial conditions were chosen either by assigning random velocities from a Gaussian 
distribution at the corresponding temperature, or starting from equal mode amplitudes 
with uniformely distributed random phases. The system is then evolved for a certain
transient time in order to start the measurements from a more generic phase-space point
as possible. In computing spectra and correlation functions a Fast Fourier Transform
routine has been used, and the data are usually averaged over an ensemble of several 
trajectories (typically between 20 and 200) to reduce statistical fluctuations.

Let us first of all comment on the dynamics of Fourier modes. The correlation of 
the fluctuating force decays on a characteristic 
time that is expected to be much shorter than the typical period of long-wavelength 
modes. Thus, if we neglect memory effects and assume that a single relaxation time 
dominates, Eq.~(\ref{langevin}) reduces to its Markovian limit (we come back to 
the discrete case): 
\be
\ddot Q_k +\gamma_k \dot Q_k + \oti_k^2 Q_k  \;=\; R_k  \quad ,
\label{markov}
\ee
where now the random force is well approximated by a Gaussian white
process
\be
\langle R_k(t)R^*_{k}(t') \rangle 
= {\gamma_k\over\beta}\delta(t-t') \quad .
\ee
For simplicity, in Eq. (\ref{markov}) we have also neglected the small frequency
shift. Eq.~(\ref{markov}) explains qualitatively the 
numerical results of Ref.\cite{alabi} where the dynamics of the modes was studied 
for the quartic FPU case. In particular, the slow diffusion of energy observed
there is immediately understood as a consequence of the fact that 
$\gamma_k/\oti_k \,\ll\,1$ for small $k$. The renormalization of the 
frequencies, that was proposed on purely phenomenological basis is, in the present 
context, a straightforward consequence of the projection approach. Furthermore, 
the effective sound velocity can now be explicitely computed by the definition 
(\ref{alfa}) (at least up to correction from Eq.~(\ref{poles})).

\begin{figure}[h]
\noindent
\centering\epsfig{figure=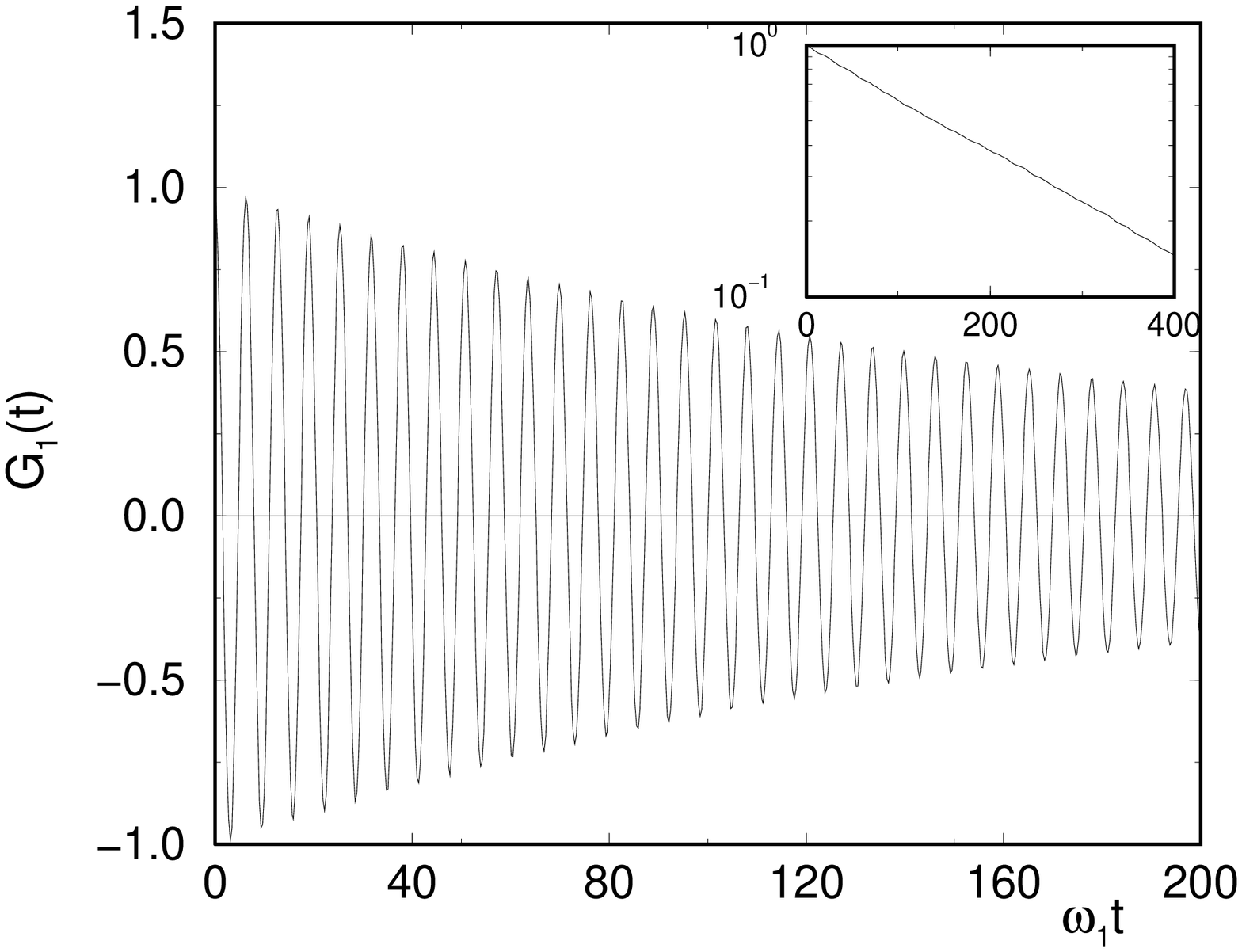,width=10cm}
\noindent
\caption{
The normalized autocorrelation ${\cal G}_1(t)$ for  
the quartic FPU model with $g_4\varepsilon=8.8$ ($T=11.07$), 
$N=256$. The inset show the autocorrelation of the fluctuations of
the mode energy $E_1$.
}
\label{autoco}
\end{figure}

\begin{figure}[h]
\noindent
\centering\epsfig{figure=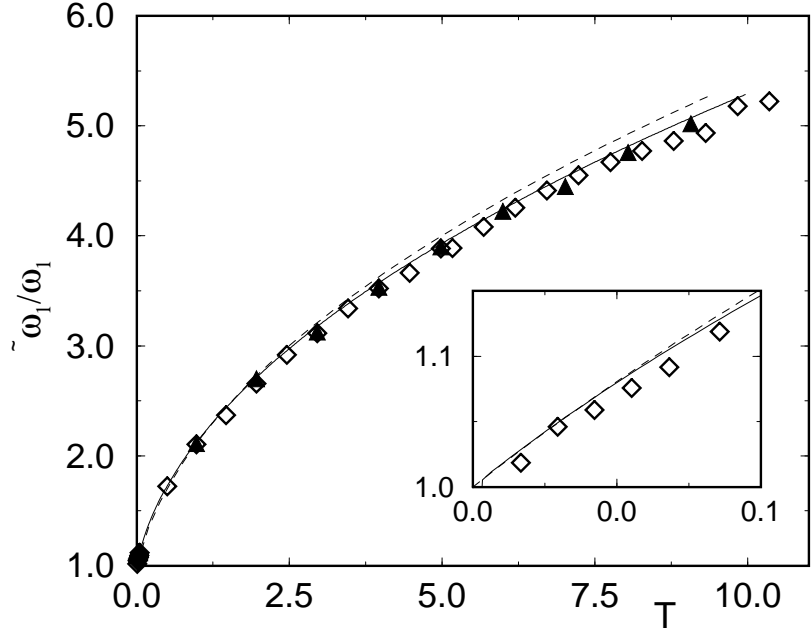,width=12cm}
\caption{
Comparison between the numerical and expected renormalized frequencies. 
The symbols are obtained measuring the oscillation period of ${\cal G}_{1}$
for $N=64,128$ (diamonds and triangles respectively).
The solid line is the theoretical value, Eq.~(\protect\ref{alfa}) and 
the dashed one is the empirical formula of Ref. [12].
The small systematic deviations are due to the small frequency shifts
expected from (\protect\ref{poles}).
}
\label{freq}
\end{figure}

At high energy ($g_4\varepsilon\gg 1$), where relaxation occours on faster time 
scales, it is relatively easier to perform direct tests of the goodness of 
Eq.~(\ref{markov}). 
We first verified that the distributions of the real and imaginary parts of $Q_k$ 
and $P_k$ are Gaussian. A typical correlation function ${\cal G}_k$ is reported 
in Fig.~\ref{autoco}. Moreover, we checked that the distribution of amplitudes 
and phase jumps agrees with what predicted from the approximation Eq.(\ref{markov}) 
\cite{strato}. For completeness, we also measured the dependence of the effective 
oscillation frequency, and compared it with the expected value given by the definition 
(\ref{alfa}) (see Fig.~\ref{freq}).

\begin{figure}[h]
\noindent
\centering\epsfig{figure=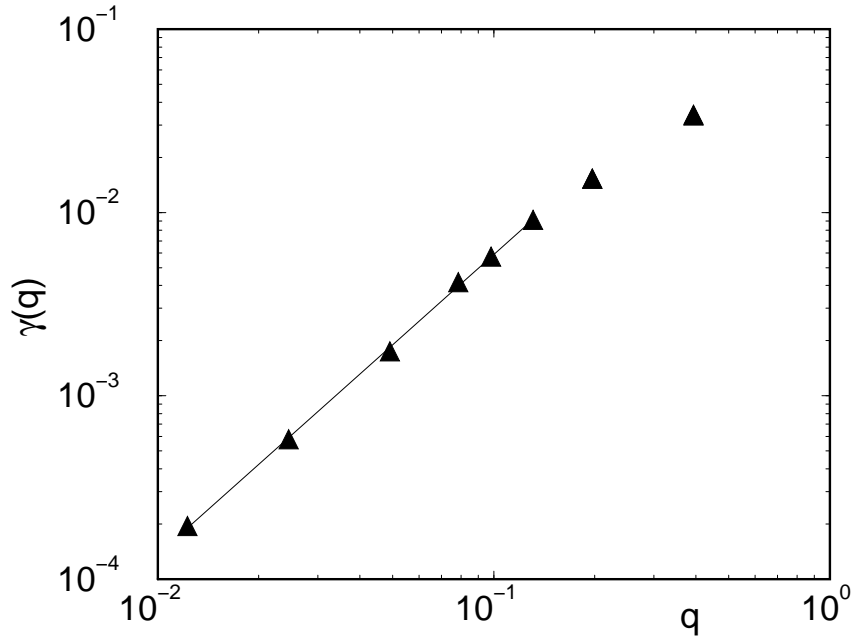,width=12cm}
\noindent
\caption{
The wavenumber dependence of the relaxation rates $\gamma(q)$ 
at $g_4\varepsilon=8.8$ for the quartic FPU potential. All the 
points were obtained from the initial 
decay of the envelope of ${\cal G}_{1}$ 
for increasing values of $N$ up to $N=2048$. The solid
line is a power-law fit $q^{1.64}$.
}
\label{ratesq}
\end{figure}

\begin{figure}[h]
\noindent
\centering\epsfig{figure=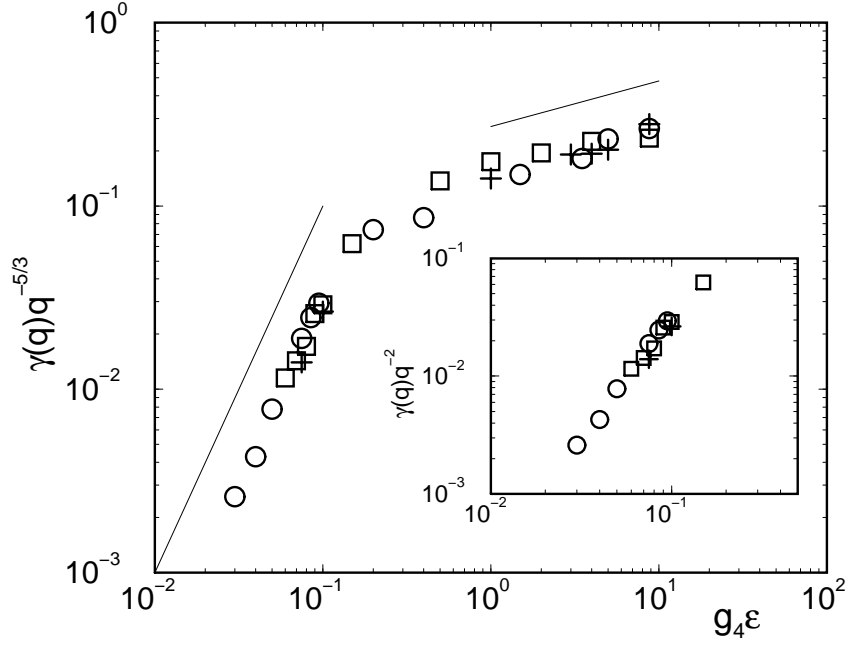,width=12cm}
\noindent
\caption{
The scaling of the relaxation rates with the energy density $\varepsilon$ 
for $N=64, 128, 256$ (circles, squares and 
pluses respectively). The inset shows the low energy part. Straight lines
correspond to $\varepsilon^2$ and $\varepsilon^{1/4}$ respectively.
}
\label{ratese}
\end{figure}

Clearly, the crucial point to be checked is the energy and wavenumber dependence of the 
$\gamma_k$. This was accomplished by measuring the initial decay of the envelope
of ${\cal G}_k$ (see  Fig.~\ref{autoco}), for several $\varepsilon$s and $k$s. 
A very good agreement with the mode-coupling prediction is 
obtained for the dependence of $\gamma_k$ on the wavenumber. The data reported in 
Fig.~\ref{ratesq} give a power law with an exponent 1.64, remarkably close to the 
expected value 5/3. Furthermore, Fig.~\ref{ratese} shows that also the 
scaling with energy is reasonably obeyed, at least within the limit of our simulations. 
Obviously, the computations become more and more time-consuming  with decreasing 
temperature due to the rapid increase of relaxation times and finite-size effects 
(see below). This imposes severe constraints on the accessible lattice lengths 
and times. 

A further analysis has been performed on the quantity
\be
{\cal E} = \sum_k E_k \quad ;\quad
E_k \;=\; {1\over 2}|P_k|^2 +{1\over 2} \oti_k^2 |Q_k|^2
\ee
which is related to the typical indicators used in equipartition studies. For
$|k|\ll N/2$ (small noise amplitudes), we expect from (\ref{markov}) that the 
effective mode energy $E_k$ is ruled by the linear equation \cite{strato}
\be
\dot E_k +\gamma_k (E_k - \langle E_k \rangle) \;=\; R_k^\prime
\ee
(see the inset of Fig.~\ref{autoco}). This means that the asymptotic behaviour of 
fluctuations $\delta{\cal E}$ of the above defined quantity are given (in the 
thermodynamic limit) by
\be
\langle \delta{\cal E}(t)\delta{\cal E}(0) \rangle \;\propto \;
{1\over N} \sum_k \, e^{-\gamma_k t} \;\approx\;
\int {dq \over 2\pi} \, e^{-\gamma(q) t} \;\propto\;
\cases{
t^{-1/2}   \quad & for $ t\ll t_C $\cr
t^{-{3/5}} \quad & for $ t\gg t_C $} \quad .
\label{crossover}
\ee
Here $t_C$ is the characteristic time scale at which a crossover between the two
relaxation behaviours occours. Although its magnitude remains undetermined, we
expect it to be very large for small temperature. The crossover should become 
actually observable in the intermediate-energy region. 

\begin{figure}[h]
\noindent
\centering\epsfig{figure=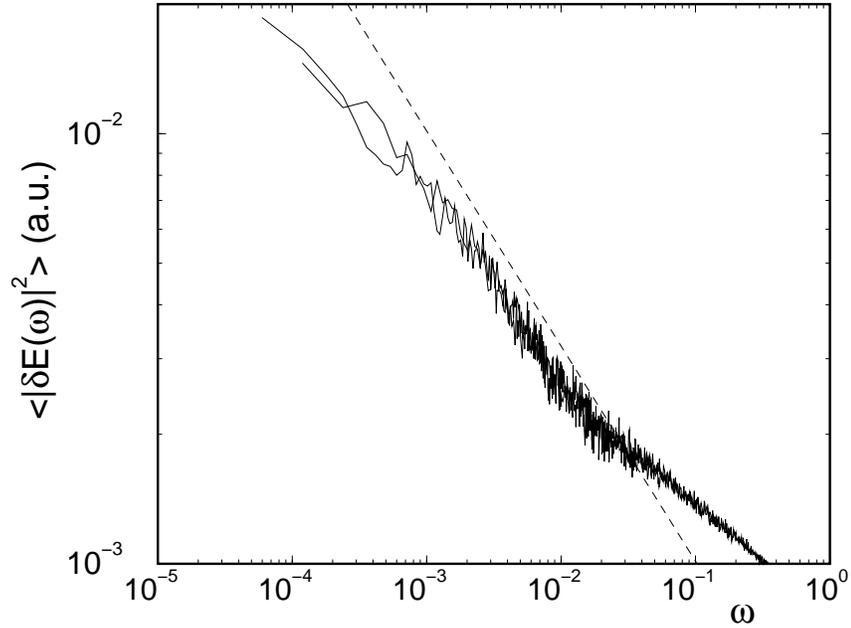,width=12cm}
\caption{
Low-frequency part of the spectrum of $\delta{\cal E}$ for the quartic 
FPU model for $g_4\varepsilon=0.05$. The solid lines
refer to simulations with $N=2048, 4096$, the dashed line
correspond to a $\omega^{-1/2}$.
A power-law fit gives an exponent -0.49$\pm$0.01 for 
$10^{-3}<\omega<10^{-2}$.
}
\label{spectra005}
\end{figure}

\begin{figure}[h]
\noindent
\centering\epsfig{figure=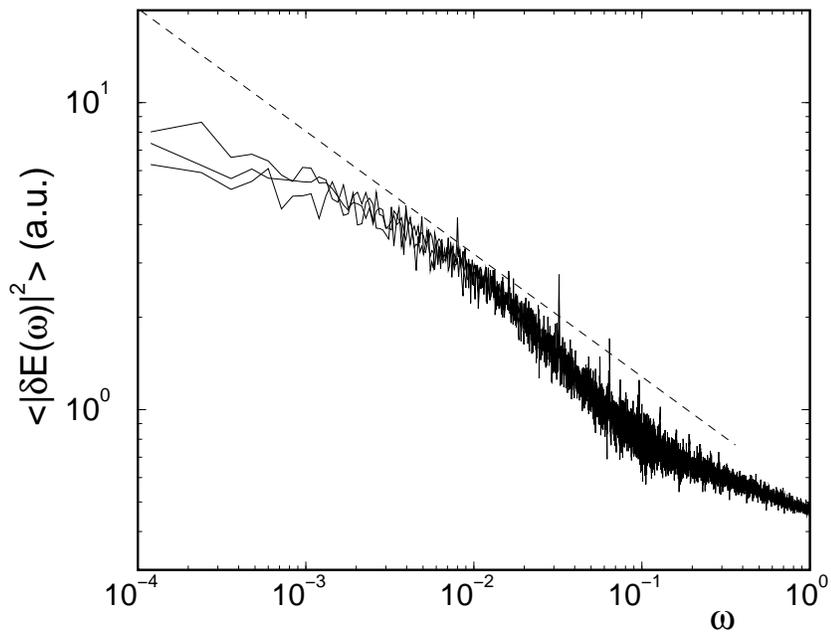,width=12cm}
\caption{
Same as Fig.~\protect\ref{spectra005} but for 
$g_4\varepsilon=0.45$. The curves refer to $N=512, 1024, 2048$ (from bottom to
top), the dashed line corresponds to the expected asymptotic law $\omega^{-2/5}$.
}
\label{spectra045}
\end{figure}

The numerical results 
are in substantial agreement with those predictions. The simulations at 
$g_4\varepsilon=0.05$ (see Fig.~\ref{spectra005}) indicate a power-law divergence 
in the spectrum of $\delta{\cal E}$ with an exponent -0.5 in the observed domain. 
The statistical accuracy and the length of the simulation are not sufficient to 
establish whether the apparent saturation for $\omega<10^{-3}$ is the beginning 
of the crossover or simply a finite-size effect. A second series of simulations 
in the intermediate energy range i.e. for $g_4\varepsilon=0.45$ is reported in 
Fig.~\ref{spectra045}. Despite the strong finite-size effects at low 
frequency, the spectra seem to approach a power-law behaviour with the 
expected mode-coupling exponent -0.4 and the data are compatible with 
(\ref{crossover}) with a value of $t_C$ of the order of 10$^2$. Since a further 
support to the validity of the mode-coupling results is also reported in \cite{noi2}, 
we reasonably conclude that no significative deviations from the theory itself 
are observed in those type of measurements.

\section{The problem of relaxation to equipartition}

At this point we want now to discuss some consequences of the above analysis on 
the so called FPU-problem. Let us first focus again on the case of a purely
quartic nonlinearity
that has been intensively (re)studied by several authors in recent years 
\cite{pettini,ruffo,thierry}. The numerical experiments have been performed by 
feeding the initial energy in a packet of modes and looking at the decay in 
time of suitable indicators of equipartition (see the quoted references
for details).

In the case when the initial excitation is around a long-wavelength mode of
wavenumber $q_*$, and the system is not too far from equipartition, we expect, 
from the discussion of Section III, that the slowest time scale 
will be of the order of 
\be 
\tau_{E} \; = \; \cases{
(g_4 \, \varepsilon \,q_*)^{-2} \quad & for $ g_4\varepsilon \ll 1 $\cr
(g_4 \, \varepsilon)^{-{1/4}} \,q_*^{-{5/3}} \quad & 
for $ g_4\varepsilon \gg 1 $}
\label{scaling} \quad.
\ee
The subscript $E$ is precisely meant to remark that it refers to the 
fluctuations close to the equilibrium state. The behaviour in the high-energy 
limit is obtained by means of Eq. (\ref{gammamct}) taking into 
account that $\varepsilon$ is roughly proportional to the temperature, as
well as the fact that $(1+\alpha)$ grows with the square root of the temperature 
itself (see Eq. (\ref{alfa})).

Indeed, the predicted regimes have been numerically observed \cite{pettini}. 
A convincing numerical evidence of scaling laws (\ref{scaling}) has been recently 
reported in Ref.~\cite{thierry}. This seem to suggest that the linear theory already 
captures the quantitative features. Moreover, qualitatively similar results were 
found for different potentials in Ref.~\cite{yoshi}. 

For what concerns the case of larger $q_*$, the simulations reported in 
Ref.~\cite{pettini} also showed that relaxation can be one or two orders of 
magnitude faster than in the previous situation. Even if the theory presented 
here applies better to the case of small $q_*$, it is worth to remark 
that its consequences are also consistent with this observation.

Obviously, the very initial stage of the approach to equilibrium may well occour
on a different time scale. In the present framework, the latter should
be interpreted as a real ``partial equilibration'' time scale $\tau_{NE}$. 
On the other hand, it seem rather reasonable that $\tau_{NE}$ will be strongly 
dependent on the chosen class of initial conditions and on the specific form
of the potential. As a matter of fact,  $\tau_{NE}$ is determined by a pure 
nonequilibrium dynamics and is of course unaccessible to the linear theory 
presented here. Despite of this,  the present results are useful to identify 
this initial stage.
As an example, we are now able to make an instructive  comparison with the 
numerical estimates of $\tau_{NE}$. In Ref.~\cite{ruffo} it is in fact found 
that, for long-wavelength excitations ($q_*\ll1$), 
$\tau_{NE} \;\sim \; N^{1/2}\, (q_*\varepsilon)^{-1}$, so that 
\be
{\tau_{NE}\over \tau_{E}} \;\sim\; \varepsilon \, N^{1/2} \,q_* \; \ll \; 1
\ee
for small $q_*$ and/or small energy. We can then conclude that the equilibration
process is, at least for this class of initial conditions, mainly 
dominated by the linear regime. In other words, we can naturally understand it
as an initial fast relaxation to a quasi-equilibrium state followed by 
the slow diffusion of energy from the long-wavelength modes. 

As already mentioned, the mechanisms determining $\tau_{NE}$ may be however rather 
complex and of very different nature depending on the initial state. An example, 
is the case of zone-boundary initial conditions ($|q_*| \approx \pi$). Those 
rapidly decay into localized chaotic excitations \cite{burla,thierry}, whose
lifetime mainly determines the time to reach the quasi-thermalized state.
Nevertheless, the comparison of our results with the numerics may indicate that 
such lifetime is, even in this case, considerably shorter than that of relaxation 
of Fourier modes \cite{thierry}. 

The high-energy scaling of Eq.~(\ref{scaling}) can be generalized to an algebraic
potential of the form $V_A(x)=g_n\;x^n$, with $n$ being an even integer. In this
case, by extending Eqs. (\ref{c3c4}) and (\ref{gammamct}), it is found that 
$\tau_E$ is proportional to $\varepsilon^{(1/n-1/2)}$.

Finally let us comment on the the FPU model with purely cubic potential ($g_4=0$). 
A recent numerical study \cite{casetti} showed a clear evidence for a divergence as 
$\varepsilon^{-3}$ of the relaxation time. Nevertheless only chains as short as $N=32$ 
were considered, and it is not completely clear if the systems is above the 
equipartition threshold \cite{shepe}. A more detailed analysis would of course be 
desirable to check the dependence of this scaling on $N$ and to compare the results 
with the ones presented here.

\section{Conclusions}

The simple mode-coupling approach provided a valuable amount of qualitative and
quantitative information on the relaxation times of the Fourier modes in an 
hamiltonian chain as (\ref{hami}). In particular, a clear physical interpretation
of them can be achieved. For the paradigmatic example of the quartic FPU model, 
the crossover between the two scaling regions at $g_4\varepsilon\approx 0.2$ 
(see again Fig.~\ref{ratese}), could be regarded as the temperature threshold 
beyond which self-consistency and ``hydrodynamic'' effects play a major role. 
Remarkably, such a scale is roughly equal to the so-called strong stochasticity 
threshold ($g_4\varepsilon\approx 0.1$) above which new dynamical effects 
(i.e. faster diffusion in phase space) are believed to appear \cite{pettini}.
From this point of view it would be challenging to try to connect the present
results with strictly dynamical properties. 

The explicit estimates given in the present work turned out to explain several 
previous results on the approach to equilibrium at finite temperature. As esemplified 
above, they are a precious information for the numerical study. For example, the 
slow temporal decay of the equipartition indicators may be better understood having
realized the existence of the long-time tails like those of Figs.~\ref{spectra005}
and \ref{spectra045}. Indeed, it is clear that the slow diffusion of energy 
has to be taken into account when studying the relaxation from an arbitrary initial 
condition. More generally, such time scales should be always considered in practice
when performing simulations with chains of large sizes. 

Although the existence of long-time tails is not surprising for a low-dimensional
system, the substantial agreement between the theory and numerical simulation is 
nonetheless a relevant result by itself. Actually, the direct verification of 
mode-coupling theories in one dimension is still current object of study, for 
example in the field of lattice gases \cite{ernst}. Furthermore, the validity of 
such theories is not granted in general. It is in fact known that they fail in 
predicting the characteristic relaxation times of spin-waves of the Heisenberg 
model in one and two dimensions \cite{reiter}.

Finally, let us remark that the theory allows to estimate the long-time tails
of the heat-flux correlation, which is directly related to the thermal conductivity
of the chain. This allows to explain quantitatively the divergence of such transport
coefficient observed in the numerical simulations \cite{noi,noi2}. 

\acknowledgements

I acknowledge useful discussions with Roberto Livi, Antonio Politi, Stefano Ruffo 
and Alessandro Torcini. Thanks also to Jochen Rau and Wolfram Just for having 
clarified to me several issues of the projection method and for the interest 
demonstrated towards this work and a careful reading of the manuscript.



\newpage

\section*{Figures}



\vspace{0.5 cm}

\vfill
\end{document}